\documentclass[final,sort&compress]{aipproc}

\layoutstyle{6x9}

\usepackage{amsmath}
\usepackage{amsfonts}
\usepackage{braket}
\usepackage{dsfont}
\usepackage{slashed}
\usepackage{graphicx}
\usepackage{hepunits}
\usepackage{empheq}

\begin{document}

\title{Higher twist parton distributions from LCWFs}

\author         {Bj\"orn Pirnay}{address={Institut f\"ur Theoretische Physik, Universit\"at Regensburg, D-93040 Regensburg, Germany} }

\begin{abstract}
 We report on the results of a recent work~\cite{Braun:2011aw}, where we explore the possibility to construct higher-twist parton distributions in a nucleon at some low reference scale from convolution integrals of the light-cone wave functions (LCWFs).
 These WFs provide one with a reasonable description of both polarized and unpolarized parton densities at large values of the Bjorken variable.
 Twist-3 parton distributions are then constructed as convolution integrals of $qqqg$ and usual three-quark WFs.
\end{abstract}

\maketitle



 Higher-twist parton distributions are conceptually very interesting as they go beyond the simple parton model description and allow one to quantify correlations between the partons.
 Unfortunately, they prove to be very elusive.
 Despite considerable efforts, very little is known even about the simplest, twist-3 distributions.
 We tried to make a step in this direction by using a representation in terms of overlap integrals of light-cone wave functions, including only Fock states with the minimum (valence) and next-to-minimum (one extra gluon) parton content.
 
 We borrow the expressions for three-quark wave functions from Ref.~\cite{Diehl:1998kh} which have been shown \cite{Diehl:1998kh,Bolz:1996sw} to provide one with a good description for quark parton densities at large $x$ and the nucleon magnetic form factor.
 The state can be described in terms of the single LCWF~\cite{Diehl:1998kh,Lepage:1980fj}
  \begin{align}\label{Ansatz}
  \ket{p,+}_{uud}&\propto \epsilon^{ijk}\int[\mathcal{D}X]_3\, \Psi_{123}^{(0)}(X)\, \Bigl(u_{i\uparrow}^\dagger(1)u_{j\downarrow}^\dagger(2)d_{k\uparrow}^\dagger(3)-u_{i\uparrow}^\dagger(1)d_{j\downarrow}^\dagger(2)u_{k\uparrow}^\dagger(3)\Bigr)\ket{0}\,.
 \end{align}
 The integration runs over all phase space configurations in which the parton plus-momentum fractions and the transverse momenta sum up to $1$ and $0$ respectively.
 For the (real) function $\Psi_{123}^{(0)}$ we adopt the simple factorized ansatz~\cite{Diehl:1998kh}
 \begin{align}\label{Psi0}
  \Psi_{123}^{(0)}&=(4\sqrt{6})^{-1}\,\phi(x_1,x_2,x_3)\, \Omega_3(a_3,x_i,k_{\perp i})\,.
 \end{align}
 The transverse momentum dependence is contained in the function $\Omega_N$, which is chosen to have a Gaussian profile and $\phi(x_i)$ is the leading-twist-3 nucleon distribution amplitude, cf.~\cite{Braun:2011aw}.
 Its normalization and shape can be estimated using QCD sum rules and lattice field theory~\cite{Chernyak:1984bm, Braun:2008ur}.
 The new contribution of our work is the inclusion of the Fock states with one additional gluon which were considered in~\cite{Diehl:1998kh} on a qualitative level.
 One can show~\cite{Braun:2011aw}, that it is sufficient to consider three independent LCWFs
 \begin{align}
  \ket{p,+}_{uudg_\downarrow} &=\epsilon^{ijk}\int[\mathcal{D}X]_4 \,\Psi^\downarrow_{1234}(X)\, g_{\downarrow}^{a,\dagger}(4)\, [t^a u_{\uparrow}(1)]_i^\dagger\, u^\dagger_{j\uparrow}(2)\,d_{k\uparrow}^\dagger(3)\ket{0}\,.\notag\\
  \ket{p,+}_{uudg^\uparrow}&=\epsilon^{ijk} \int[\mathcal{D}X]_4\Big\{\Psi^{\uparrow(1)}_{1234}(X)\,[t^au_{\downarrow}(1)]_i^\dagger\Bigl(u_{j\uparrow}^\dagger(2)d_{k\downarrow}^\dagger(3)-d_{j\uparrow}^\dagger(2)u_{k\downarrow}^\dagger(3)\Bigr)g_{\uparrow}^{a,\dagger}(4)
  \notag\\
  &+\Psi^{\uparrow(2)}_{1234}(X)u^\dagger_{i\downarrow}(1)\Big( [t^a u_{\downarrow}(2)]^\dagger_j\, d^\dagger_{k\uparrow}(3)-[t^a d_{\downarrow}(2)]^\dagger_j\, u^\dagger_{k\uparrow}(3)\Big)g_{\uparrow}^{a,\dagger}(4)\Big\}\ket{0}\,.  \label{gwf}
 \end{align}
 For the transverse momenta we consider an analogous, 4-variable, ansatz to Eq.~\eqref{Psi0} and adopt the normalization of these new wave functions from a QCD sum rule approach~\cite{Braun:2011aw}.


 Having specified the wave functions, we calculate the quark and gluon polarized and unpolarized parton distributions.
 \begin{figure*}[t]
  \includegraphics[width=0.4\linewidth,clip=true]{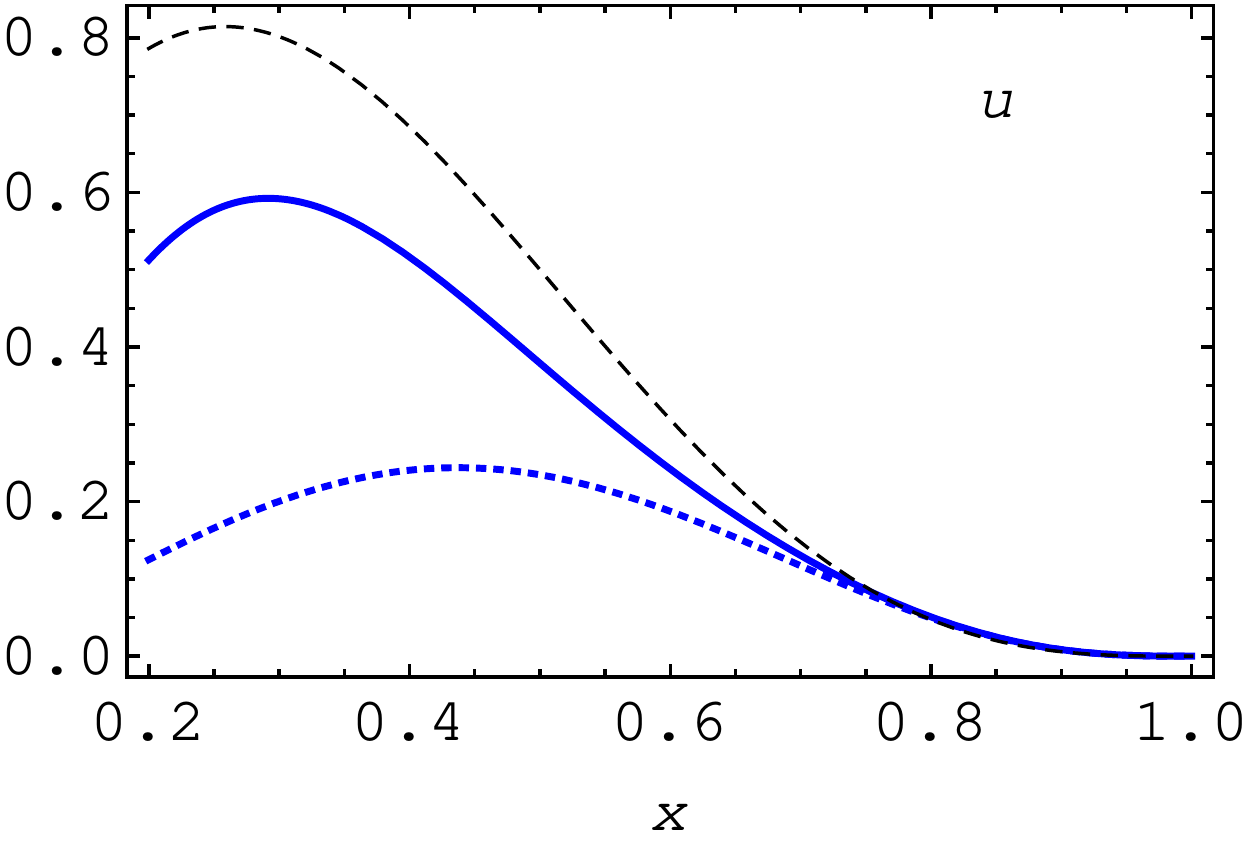} \includegraphics[width=0.4\linewidth,clip=true]{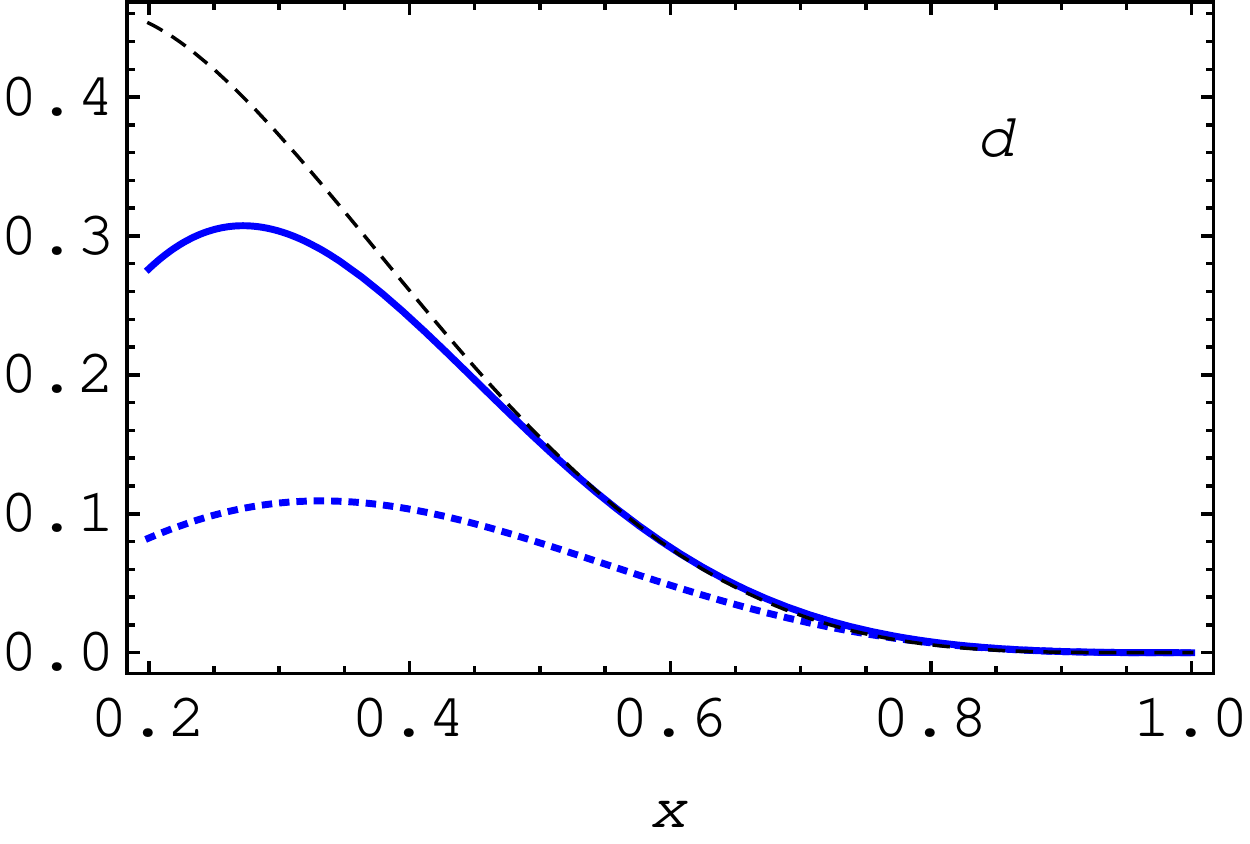}
  \caption{\footnotesize Up- and down quark distributions, $xu(x)$ and $xd(x)$. The black (short dashed) curves correspond to the GRV parametrizations~\cite{Gluck:1998xa} at the scale $\mu^2=1GeV^2$.
  The solid blue curve is our model prediction taking, the contribution of the valence state alone is shown by dots for comparison.
  }
  \label{fig:PD}
 \end{figure*}
 As an example, the results for the up- and down-quarks are shown in Fig.~\ref{fig:PD}.
 It is seen that this simple approximation captures main features of parton distributions at large $x$ surprisingly well, although more sophisticated models are certainly needed for a better quantitative description.


 A description of twist-3 observables in the framework of collinear factorization involves quark-antiquark-gluon correlation functions which are defined as matrix elements of nonlocal (light-ray) three-particle operators.
 All matrix elements in question can be expressed in terms of two correlation functions $\mathcal{Q}^{\uparrow(\downarrow)}_q(x)$ defined as
 \begin{align}\label{qpm1}
  {}_{uud}\braket{p,+|\bar q^{\uparrow}_+(z_3)f_{++}(z_2)q^{\uparrow}_+(z_1)|p,-}_{uudg^\uparrow}&=-2ip_+^2\int \mathcal{D}x\, e^{-ip_+\sum x_iz_i} \mathcal{Q}^{\uparrow}_q(x)\,,\notag\\
  {}_{uud}\braket{p,+|\bar q^{\downarrow}_+(z_1)f_{++}(z_2)q^{\downarrow}_+(z_3)|p,-}_{uudg^\uparrow}&=-2ip_+^2\int \mathcal{D}x\, e^{-ip_+\sum x_iz_i} \mathcal{Q}^{\downarrow}_q(x)\,,
 \end{align}
 where the subscript $q=u,d$ stands for quark flavor.
 Here and below the integration measure is defined as $\int\mathcal{D}x=\int dx_1 dx_2 dx_3 \,\delta(\sum x_i)$.

 
 The structure function $g_2(x_B, Q^2)$ is given by the sum of the Wandzura-Wilczek (WW) and genuine twist-3 contributions
 \begin{align}
  g_2(x_B,Q^2) = g^{WW}_2(x_B,Q^2)+g^{tw-3}_2(x_B,Q^2)\,.
 \end{align}
 The twist-3 contribution $g^{tw-3}_2(x_B,Q^2)$ can be written as
 \begin{align}
  g_2^{\mathrm{tw-3}}(x_B,Q^2)=\sum_{q}\frac{ge_q^2}{2m_N}\int\mathcal Dx\,\mathcal{Q}^+_q(\boldsymbol x)&\biggl[\biggl(\frac{1}{x_3}+\frac{P_{13}-1}{x_2}\biggr)\frac{\theta(x_3-x_B)}{x_2x_3}-\frac{\delta(x_3-x_B)}{x_2x_3}\biggr] \,,\nonumber
 \end{align}
 where $\mathcal{Q}^+_q(x)=\mathcal{Q}^\downarrow_q(x)+\mathcal{Q}^\uparrow_q(x) + (x\leftrightarrow -x )$.
  \begin{figure*}[t]
  \includegraphics[width=0.4\linewidth,clip=true]{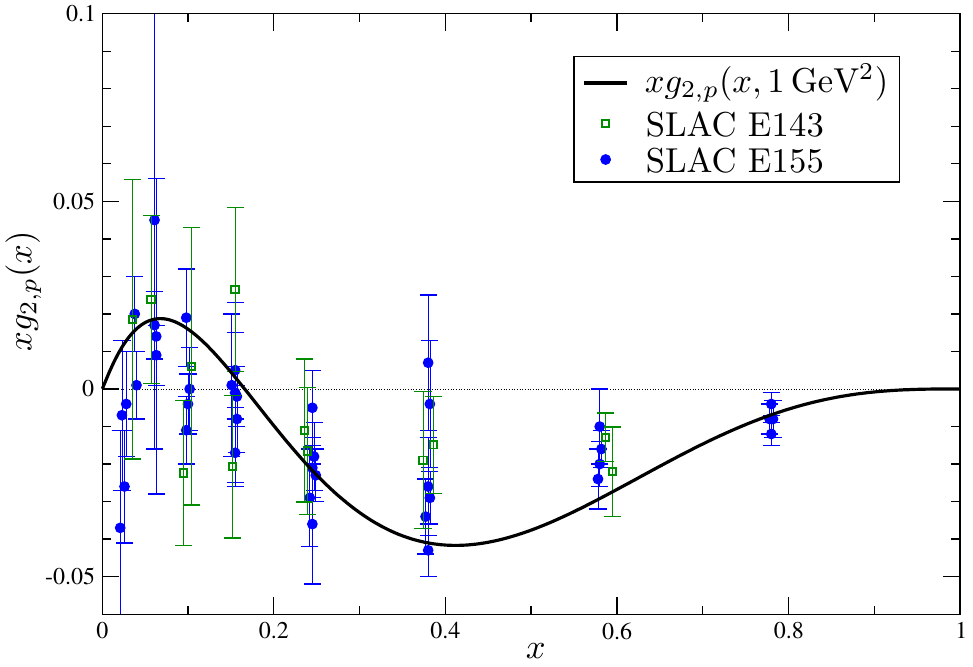}
  \includegraphics[width=0.4\linewidth,clip=true]{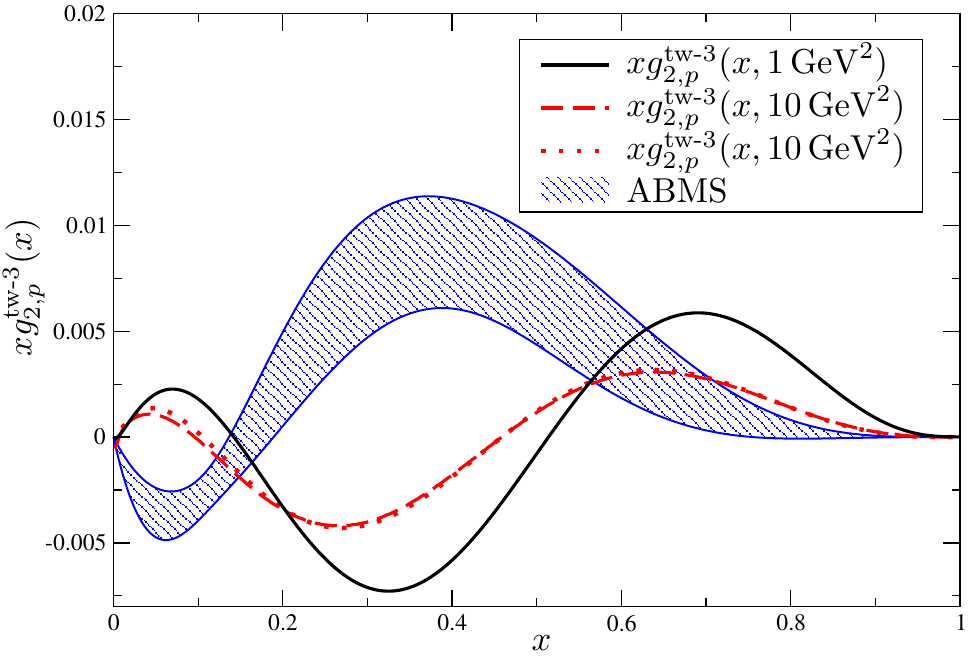}
  \caption{\footnotesize
   Left: Experimental results on the proton structure function $g_2(x_B,Q^2)$ compared to our model calculation at the scale $Q^2=1$~GeV$^2$.
   Right: The twist-3 contributions  $x g^{tw-3}_2(x_B,Q^2)$ for the proton compared to the analysis in Ref.~\cite{Accardi:2009au} (shaded areas).
   Our model predictions at the scale $Q^2=1$~GeV$^2$ and $Q^2=10$~GeV$^2$ are shown by the black solid and dashed red curves, respectively.
  }
  \label{fig:g2}
 \end{figure*}
 Our results for the full structure function $g_{2}(x_B,Q^2)$ are compared to the experimental data in Fig.~\ref{fig:g2} (left) and, separately, for the twist-3 contribution to the analysis in Ref.~\cite{Accardi:2009au} (right).
 The twist-3 contributions are shown at the model scale $Q^2=1$~GeV$^2$ and after the evolution to a higher scale $Q^2=10$~GeV$^2$, cf. \cite{Braun:2009mi} and references therein.


 The quark-antiquark-gluon correlation functions considered here are precisely those responsible for transverse single spin asymmetries (SSA) observed in different hadronic reactions.
 The distribution $T_{\bar q F q}({x})$ introduced in this context in Ref.~\cite{Braun:2009mi} is expressed in terms of $\mathcal{Q}_q^{\uparrow(\downarrow)}$--functions as follows:
 \begin{align}
  T_{\bar q F q}(x) &= -\tfrac{g}2\big[P_{13}\mathcal{Q}_q^{\uparrow}(x)+\mathcal{Q}_q^{\uparrow}(-x) +\mathcal{Q}_q^{\downarrow}(x)+P_{13}\mathcal{Q}_q^{\downarrow}(-x)\big]\,,
 \end{align}
 \begin{figure*}[t]
  \includegraphics[width=0.85\linewidth,clip=true]{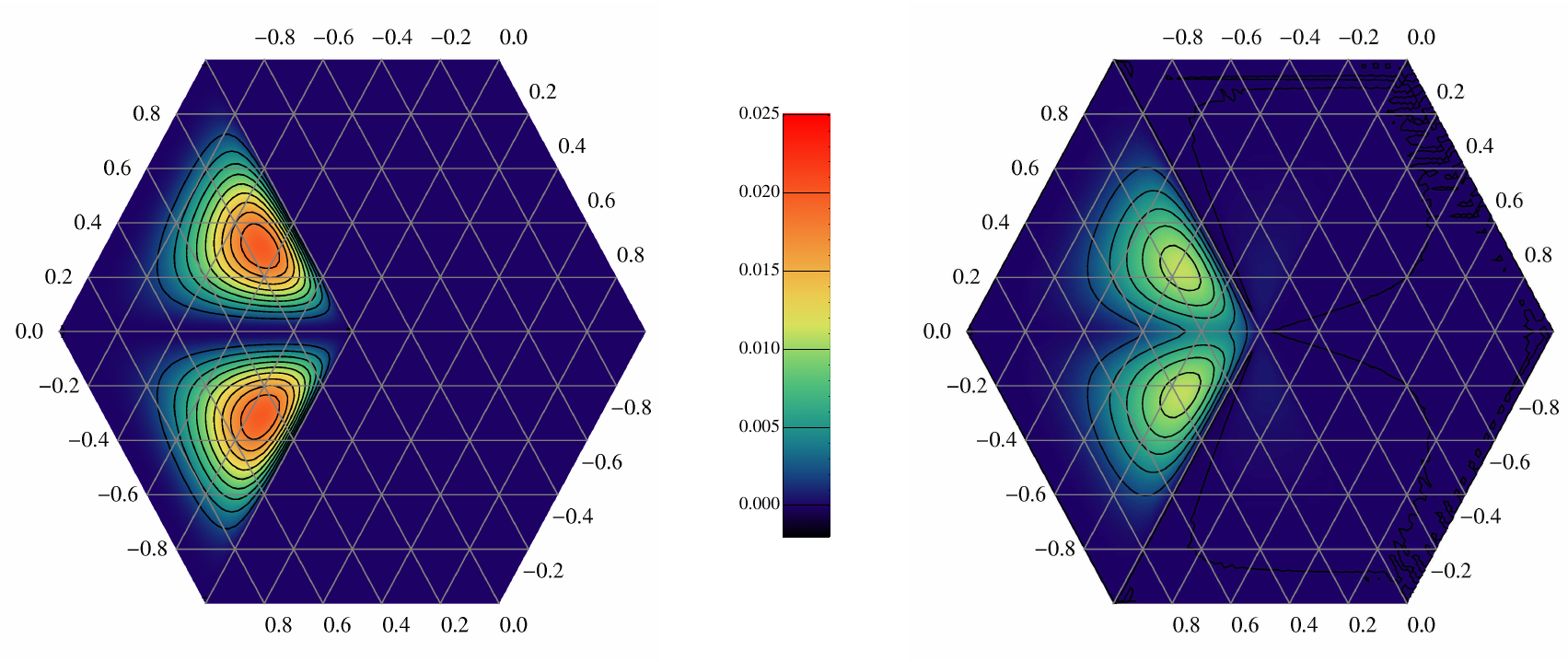}
  \caption{\footnotesize
   The quark-antiquark-gluon twist-3 correlation function $-T_{\bar d F d}(x)$
   at the reference scale $\mu^2=1$~GeV$^2$ (left) and $\mu^2=10$~GeV$^2$ (right).
   }
  \label{fig:TdFd}
 \end{figure*}
 In the framework of collinear factorization, SSAs originate from imaginary (pole) parts of propagators in the hard coefficient functions.
 In the leading order, taking a pole part enforces vanishing of one of the momentum fractions in the twist-3 parton distribution, and are classified as soft gluon pole (SGP) or soft fermion pole (SFP).
 Such ``pole'' contributions are therefore considered to be main source of the observed asymmetries and can be estimated from the available experimental data~\cite{Kouvaris:2006zy,Kanazawa:2010au}.

 For definiteness we show the results for the down quark correlators, $q=d$.
 Since our approximation for the nucleon wave function does not contain antiquarks, the $T_{\bar d F d}$ distribution is nonzero in two triangular regions of its hexagonal support only, cf. Fig.~\ref{fig:TdFd}.
 Moreover, it vanishes at the boundaries of parton regions where one of the momentum fractions goes to zero, and, hence, both SGP and SFP terms vanish as well.
 This property is an obvious artefact of the truncation of the Fock expansion to a few lowest components:
 The LCWF of each Fock state vanishes whenever momentum fraction of any parton goes to zero and the same property holds true for the correlation functions.
 It is easy to see that both the SGP and SFP contributions reappear once QCD evolution is taken into account~\cite{Braun:2009mi}.

 
 Since our approximation for the nucleon wave function only includes a few lowest Fock components, and since the LCWF of each Fock state vanishes whenever momentum fraction of any parton goes to zero, both SGP and SFP terms vanish at the scale where the model is formulated.
 They are, however, generated by QCD evolution that brings in multiple soft gluon emission.
 Our results suggest that realistic dynamical models of the the twist-3 distributions can be obtained following the GRV-like approach on the level of WFs, i.e.  assuming that the nucleon state at a very low scale can be described in terms of a few Fock components, including the valence quarks, one additional gluon and, probably, a quark-antiquark pair, and applying QCD evolution equations.


\begin{theacknowledgments}
 The author would like to thank V.M. Braun and A.N. Manashov for collaboration and fruitful discussions. Many thanks go to the organizers of the workshop and the convenors of the Spin Physics working group.
\end{theacknowledgments}

\bibliographystyle{aipproc}

\end{document}